\documentclass[sigconf,natbib]{acmart}
\usepackage{cleveref}

\usepackage{bm}
\usepackage{multirow}
\usepackage[acronym]{glossaries}
\makeglossaries

\newacronym{ai}{AI}{Artificial Intelligence}
\newacronym{auc}{AUC}{Area Under the Curve}
\newacronym{dl}{DL}{Deep Learning}
\newacronym{gdpr}{GDPR}{General Data Protection Regulation}
\newacronym{gnn}{GNN}{Graph Neural Network}
\newacronym{lstm}{LSTM}{Long Short-Term Memory}
\newacronym{ml}{ML}{Machine Learning}
\newacronym{mlp}{MLP}{Multi-Layer Perceptron}
\newacronym{mse}{MSE}{Mean Squared Error}
\newacronym{ndcg}{NDCG}{Normalized Discounted Cumulative Gain}
\newacronym{relu}{ReLU}{rectified linear unit}
\newacronym{seq2seq}{Seq2Seq}{sequence-to-sequence}
\newacronym[plural={SERPs},firstplural={Search Engine Result Pages}]{serp}{SERP}{Search Engine Results Page}
\newacronym{sota}{SOTA}{state of the art}
\newacronym{vit}{ViT}{Vision Transformer}
\newacronym{seo}{SEO}{Search Engine Optimisation}
\newacronym{ppa}{PPA}{pay-per-attention}
\newacronym{tft}{TFT}{Total Fixation Time}
\newacronym{tfc}{TFC}{Total Fixation Count}
\newacronym{aoi}{AOI}{Areas of Interest}
\newacronym{www}{WWW}{World Wide Web}
\newacronym{ir}{IR}{Information Retrieval}
\newacronym{dom}{DOM}{Document Object Model}

\usepackage{booktabs}
\usepackage{adjustbox}
\usepackage{soul}
\usepackage{amsmath}

\usepackage{microtype}
\usepackage{balance}
\usepackage{xcolor}
\usepackage{tabularx}
\usepackage{colortbl}
\usepackage{paralist}
\usepackage{array}
\usepackage{graphicx, subfig}
\usepackage{tcolorbox}

\newcolumntype{L}[1]{>{\raggedright\let\newline\\\arraybackslash\hspace{0pt}}m{#1}}
\newcolumntype{C}[1]{>{\centering\let\newline\\\arraybackslash\hspace{0pt}}m{#1}}
\newcolumntype{R}[1]{>{\raggedleft\let\newline\\\arraybackslash\hspace{0pt}}m{#1}}

\usepackage[inline]{enumitem}

\definecolor{direct-right_ad_colour}{RGB}{165, 89, 170}
\definecolor{organic_ad_on_bottom_colour}{RGB}{89, 168, 156}
\definecolor{organic_ad_on_top_colour}{RGB}{240, 197, 113}
\definecolor{direct-top_ad_colour}{RGB}{224, 43, 53}

\AtBeginDocument{
  }

\copyrightyear{2025}
\acmYear{2025}
\setcopyright{cc}
\setcctype{by}
\acmConference[SIGIR '25]{Proceedings of the 48th International ACM SIGIR Conference on Research and Development in Information Retrieval}{July 13--18, 2025}{Padua, Italy}
\acmBooktitle{Proceedings of the 48th International ACM SIGIR Conference on Research and Development in Information Retrieval (SIGIR '25), July 13--18, 2025, Padua, Italy}
\acmDOI{10.1145/3726302.3729891}
\acmISBN{979-8-4007-1592-1/2025/07}

\begin{document}

\title{AdSight: Scalable and Accurate Quantification of User Attention in Multi-Slot Sponsored Search}

    \author{Mario Villaizán-Vallelado}
	\orcid{0009-0002-0754-1742}
	\affiliation{
		\institution{Universidad de Valladolid}
		\institution{Telef\'{o}nica Scientific Research}
		\country{Spain}
	}

    \author{Matteo Salvatori}
	\orcid{0000-0003-1499-6024}
	\affiliation{
		\institution{Telef\'{o}nica Scientific Research}
		\country{Spain}
        }

    \author{Kayhan Latifzadeh}
	  \orcid{0000-0001-6172-0560}
	\affiliation{
		\institution{University of Luxembourg}
		\country{Luxembourg}
        }

    \author{Antonio Penta}
	\orcid{0000-0003-3796-8245}
	\affiliation{
		\institution{ICREA-Universitat Pompeu Fabra}
        \institution{Barcelona School of Economics}
		\country{Spain}
        }

    \author{Luis A. Leiva}
	\orcid{0000-0002-5011-1847}
	\affiliation{
		\institution{University of Luxembourg}
		\country{Luxembourg}
        }

    \author{Ioannis Arapakis}
	\orcid{0000-0003-2528-6597}
	\affiliation{
		\institution{Telef\'{o}nica Scientific Research}
		\country{Spain}
        }

\renewcommand{\shortauthors}{Mario Villaizán-Vallelado et al.}

\begin{abstract}
Modern \glspl{serp} present complex layouts where multiple elements compete for visibility. 
Attention modelling is crucial for optimising web design and computational advertising, 
whereas attention metrics can inform ad placement and revenue strategies. 
We introduce AdSight, a method leveraging mouse cursor trajectories 
to quantify in a scalable and accurate manner user attention in multi-slot environments like \glspl{serp}. 
AdSight uses a novel Transformer-based \acrlong{seq2seq} architecture where the encoder processes cursor trajectory embeddings, 
and the decoder incorporates slot-specific features, enabling robust attention prediction across various \gls{serp} layouts. 
We evaluate our approach on two \acrlong{ml} tasks: 
(1)~\emph{regression}, to predict fixation times and counts; 
and (2)~\emph{classification}, to determine some slot types were noticed. 
Our findings demonstrate the model's ability to predict attention with unprecedented precision, 
offering actionable insights for researchers and practitioners.
\end{abstract}



\keywords{SERP; Sponsored search; Online advertising; Mouse cursor tracking; Multi-slot ads; Direct displays; User attention; Neural networks}

\received{20 February 2024}
\received[revised]{12 March 2025}
\received[accepted]{5 June 2025}

\maketitle

\section{Introduction\label{sec:introduction}}

The evolution of the Web has had a significant impact on how saliency and attention are both understood and computed.
Initially, the Web consisted primarily of static HTML pages featuring mostly text and basic hyperlinks. 
Saliency was mainly determined by textual content, ignoring the layout. 
Over time, the incorporation of visual elements like images, logos, banners, and basic animations increased the complexity of web layouts. 
Thus, saliency considerations expanded to include user attention patterns.
The Web 2.0 brought another paradigm shift, introducing dynamic and user-generated content. 
Social media platforms emerged and websites became further convoluted in layout and functionality. 
This evolution increased the number of elements competing for user attention, 
making its computation more crucial for effective web design.

Currently, new challenges and opportunities arise, such as:
\begin{enumerate*}[label={(\arabic*)}]
\item the increased use of video, audio, and interactive elements, creating more complex visual hierarchies; 
\item the adoption of responsive designs, requiring considering diverse screen sizes and devices; and 
\item the variability in user attention patterns, even when viewing the same visual stimuli. 
\end{enumerate*}
The interplay between cognitive processes, such as reading, and the competition among visual elements 
further compounds the problem of attention prediction in environments with diverse layouts~\cite{Chakraborty2023}.

These challenges have implications for search engines and social media, 
where advertising revenue funds many free online services. 
For example, ads generate around one third of the advertising revenue for major search engines,
totalling approximately \$50 billion annually~\cite{Gleason2024}. 
Modern \glspl{serp} incorporate elements such as snippets, one-box answers, and side panels, 
creating non-trivial patterns of user attention. 
Such configurations can result in ``good abandonments'', where users find information without clicking, 
complicating the assessment of user satisfaction~\cite{Chuklin2016} or display noticeability~\cite{Bruckner2021When}.

Accurate attention prediction is essential for optimizing computational advertising. 
It enables strategic ad placements that enhance visibility and reduce intrusive experiences, 
potentially increasing ad effectiveness. 
Furthermore, as search engines adopt conversational interfaces, 
opportunities emerge for integrating native advertising within generative search results, 
blending product placements with organic responses~\cite{Zelch24}. 
Attention prediction also offers significant potential for advancing digital advertising models, 
such as the \gls{ppa} auction scheme~\cite{arapakis2020}, 
which considers attention likelihood rather than mere clicks or ad impressions. 
While diagnostic technologies~\cite{Arapakis20_mtdl} 
can improve our understanding of user engagement and inform smarter auction models, 
existing solutions remain limited, as they consider single ads in isolation.
We address these key challenges with the following \textbf{contributions}:

\begin{itemize}
    \item A novel and scalable method to quantify user attention 
    from mouse movements in multi-slot environments like \glspl{serp}. 

    \item A Transformer-based \gls{seq2seq} model, 
    where the encoder processes cursor trajectory embeddings 
    and the decoder incorporates slot-specific features. 

    \item Demonstration of the utility of AdSight as a scalable and accurate attention measurement method 
    through two \gls{ml} tasks, namely (1)~\emph{Regression}: Predicting fixation times and counts with near-second precision; 
    and (2)~\emph{Classification}: Identifying whether a user has noticed some advertisements. 
    Both tasks utilise ground truth from organic eye tracking data.

    \item Exploration of cursor data representations and embeddings, 
    showing how auxiliary slots can serve as additional \glspl{aoi} to enhance model learning.  
\end{itemize}

\section{Related work\label{sec:Related_works}}

The analysis of mouse cursor trajectories and their temporal dynamics 
have offered powerful insights into the underlying cognitive processes 
of many everyday tasks such as e-shopping~\cite{Gwizdka22}. 
Actually, the first applications of mouse tracking 
can be traced to web browsing analysis~\cite{Mueller01_cheese, Chen:2001:MCT:634067.634234}.
Experimental psychology and cognitive science research
have also tapped into mouse movement analysis to study for example 
consumer's choice~\cite{Soman98, Zauberman03}, 
decision-making processes~\cite{Maldonado19, Stillman18}, 
and motor control performance~\cite{Smith1999}.

Historically, in \gls{ir}, eye tracking has been used to study user behaviour~\cite{Goldberg:2002:ETW:507072.507082, Arapakis:2014:UEO:3151365.3151368, Li:2017:TMI:3041021.3054182, Brightfish18, Papoutsaki16}.
However, eye tracking requires specialized equipment and is typically confined to lab settings.
On the contrary, mouse tracking requires no dedicated equipment or extensive training 
and can be deployed online, in a cost-effective and scalable manner, while users browse within their natural environment~\cite{Leiva:2013:WBB:2540635.2529996}. 
Further, mouse movements are considered as a reasonable proxy for user's gaze, 
especially on \glspl{serp}~\cite{Huang:2011:NCN:1978942.1979125, Huang:2012:USU:2207676.2208591, Arapakis:2016:PUE:2911451.2911505, leiva2020attentive, Guo:2012:BDT:2187836.2187914, Speicher:2013:TPR:2505515.2505703},
highlighting mouse tracking as scalable alternative to eye tracking~\cite{Huang:2011:NCN:1978942.1979125, Guo11_success, Hassan10_success}.
Applications of mouse tracking in \gls{ir} include:
inform usability tests~\cite{Atterer:2006:KUM:1135777.1135811},
predict user engagement~\cite{Arapakis:2014:UWE:2661829.2661909}
and intent~\cite{Guo:2008:EMM:1390334.1390462, MARTINALBO2016989},
detect searcher frustration~\cite{Feild:2010:PSF:1835449.1835458}
and page abandonment~\cite{Diriye:2012:LSS:2396761.2398399, Bruckner2021When},
or identify open-ended behaviours~\cite{Jaiswal23}.

\subsection{User Modelling}

Tracking where on the page a user's mouse cursor hovers or clicks provides a surrogate signal for gaze fixation, revealing the focus of attention, which can be used to learn the users' latent interests.

\subsubsection{Inferring User Interest}
\label{sec:interest}

Early work considered simple, coarse-grained features derived from mouse cursor data
to be reasonable measurements of user interest~\cite{Claypool:2001:III:359784.359836, Shapira:2006:SUK:1141277.1141542}.
Follow-up research transitioned to more fine-grained mouse cursor features~\cite{Guo:2008:EMM:1390334.1390462, Guo:2010:RBJ:1835449.1835473}
that were shown to be more effective.
These approaches have been directed at predicting open-ended tasks
like search success~\cite{Guo:2012:PWS:2396761.2398570} 
or search satisfaction~\cite{Liu:2015:DUD:2766462.2767721}.
In a similar vein, Huang et al.~\cite{Huang:2012:ISM:2348283.2348313, Huang:2011:NCN:1978942.1979125}
modeled mouse cursor interactions and extended click models 
to compute more accurate relevance judgements of search results.

\subsubsection{Inferring Visual Attention}
\label{sec:attention}

Mouse cursor tracking has been also used to survey the visual focus of users,
thus revealing valuable information
regarding the distribution of user attention over the various \glspl{serp} components.
Despite the technical challenges that arise from this analysis,
previous work has shown the utility of mouse movement patterns
to predict reading~\cite{Hauger11} or hesitations~\cite{Mueller01_cheese}.
\citet{Lagun:2014:DCM:2556195.2556265} introduced the concept of motifs,
or frequent cursor subsequences, in the estimation of search result relevance.
Similarly, \citet{Liu:2015:DUD:2766462.2767721} applied the motifs concept to \glspl{serp}
and predicted search result utility, searcher effort, and satisfaction at a search task level.
\citet{Boi2016} proposed a method for predicting whether the user is looking at the content pointed by the cursor,
exploiting the mouse cursor data and a segmentation of the contents in a web page. 
Lastly, Arapakis et al.~\cite{Arapakis:2016:PUE:2911451.2911505, arapakis2020} investigated user engagement with direct displays on \glspl{serp}, supporting the utility of mouse cursor data for measuring display-level user attention. They proposed a pay-per-attention scheme, where advertisers are charged only when users noticed the ads shown on \glspl{serp}.

\subsubsection{Inferring Personally Identifiable Information}
\label{sec:emotion}

The link between mouse cursor movements and psychological states has been researched since the early 90s~\cite{Accot1997, Card:1987}. Some studies explored the utility of mouse cursor data for predicting the user's emotional state~\cite{Zimmermann2003, Kaklauskas2009}, including frustration~\cite{Kapoor:2007, Azcarraga:2012} and anxiety~\cite{Yamauchi:2013}. This led to profiling applications, such as decoding age and gender~\cite{Yamauchi2014, Kratky:2016, Pentel2017}, highlighting privacy concerns from unethical and unregulated mouse cursor tracking~\cite{Leiva2021MyMouse}.

\subsection{Sponsored Search}
\label{sec:ppa}

Commercial search engines have been interested in how users interact with \glspl{serp},
to anticipate better ad placement in sponsored search or optimize the page layout. 
Earlier models of scanning behaviour in \glspl{serp} were assumed to be linear, 
as users tend to explore the search results from top to bottom. 
Today this is no longer the case, 
since \glspl{serp} now include several heterogeneous modules (direct displays) 
like image and video search results, featured snippets, or feature-rich ads~\cite{Arapakis:2015:KYO:2806416.2806591, Roy22_serps, Shao22}.
To account for this heterogeneity, 
\citet{Diaz13} proposed a generalization of the classic linear scanning model 
which incorporated ancillary page modules, 
whereas \citet{Huang:2012:ISM:2348283.2348313} and~\citet{Speicher:2013:TPR:2505515.2505703} 
modelled mouse cursor interactions
by extending click models to compute more accurate relevance judgements of search results. 
\citet{Liu:2014:SRT:2661829.2661907} noticed that 
almost half of the search results fixated by users are not being read, 
since there is often a preceding skimming step
in which users quickly look at the page structure.
Based on this observation, they propose a two-stage examination model:
a first ``from skimming to reading'' stage and a second ``from reading to clicking'' stage.
Interestingly, they showed that both stages can be predicted from mouse cursor movements.

The closest work to ours is by Arapakis et al.~\cite{Arapakis:2016:PUE:2911451.2911505, Arapakis20_mtdl, arapakis2020}, who investigated user engagement with direct displays on \glspl{serp}, one element (slot) at a time. 
Similarly, we implement a diagnostic technology for measuring user attention to slots. 
However, our work differs in two key aspects. 
First, we propose a novel predictive modelling framework based on Transformers for detecting attention to \textit{multiple slots}, 
including organic and direct-display ads, in various positions, within Google \glspl{serp}. 
Second, we use eye tracking to collect \emph{objective} ground-truth labels of user attention, 
weighing \glspl{aoi} in \glspl{serp} for enhanced mouse movement analysis. 
To our knowledge, we are the first to introduce this methodology.

\section{User study\label{sec:user_study}}

\subsection{Participants}
A total of 47 participants (20\,F, 25\,M), aged 19--44 years ($M=29.66$, $SD=6.46$) 
were recruited via mailing lists and provided written consent before the experiment. 
All participants had normal or corrected-to-normal vision 
and self-reported being proficient in English.
Compensation was 20\,EUR.

\subsection{Design}
We designed a between-subjects experiment with two independent variables: 
(1)~slot format (organic and direct-display ads)
and (2)~slot position (direct-display ads placed on the top-left or top-right part, 
and organic ads placed at the top or bottom). 
The dependent variable was user attention, measured via gaze fixations.

\subsection{Procedure}
The experiment was divided into eight blocks, each comprising 10 trials.
The first two blocks (not considered for analysis) 
were used as warm-up tasks to help participants familiarize themselves with the procedure.
Each trial comprised an independent transactional task.
The eye-tracker was recalibrated before the start of every block.
Participants were given a rest period of at least one minute between blocks. 
At the beginning of each trial, participants were presented with a product title and an associated query text. 
They were instructed to imagine they were planning to purchase that product 
and then navigated to the \glspl{serp}. 
They had up to one minute to review the results and click on whatever item they would typically choose in those circumstances. 
After making their selection, participants were asked to confirm their choice. 
If they chose not to confirm, they could continue browsing for another minute.

\subsection{Apparatus}

\noindent\textbf{Equipment.} We used a 17-inch Dell 1707FP 1280x1024\,px LCD monitor with a refresh rate of 60\,Hz. 
All \glspl{serp} were presented in full-screen mode using Google Chrome browser. 
Eye-tracking data was collected using a Gazepoint GP3 HD eye tracker, operating at a sampling rate of 150\,Hz. 
Mouse movements were captured with a Dell MS116 mouse.

\noindent\textbf{Search Queries.} We extracted product titles from the Amazon Product Reviews dataset~\cite{McAuley15_dataset}
and used them as queries on Google search by prepending ``buy'' to each title. 
This yielded 4,410 distinct \glspl{serp} in English, 
featuring both organic and direct display ads. 
The \glspl{serp} were randomly organised into 10 blocks,
282 of which were eventually included in the study.

\noindent\textbf{\gls{serp} Layout.} Organic ads appeared either at the top
or bottom 
parts of the \glspl{serp}, whereas direct display ads appeared in the upper-left
or upper-right
parts. 
Each \gls{serp} had either a left-aligned or a right-aligned direct display ad. 
It could also include organic ads or consist only of organic ads, without any direct display ads.

\noindent\textbf{Dataset.} Our final dataset 
consisted of 2,776 trials,\footnote{We excluded 44 trials due to malformed log files.} 
with each trial having four associated files: 
(1)~a rendered full-page screenshot of the corresponding \gls{serp}; 
(2)~a `slot boundaries' file containing the bounding box coordinates of ad blocks on the screenshot, 
(3)~an `eye fixations' file in the format $(t, x, y, d)$, 
where $t$ represents timestamps, 
$x$ and $y$ denote gaze positions (relative to the top-left corner of the screen), 
and $d$ indicates fixation duration; 
(4)~a `mouse movements' file in the format $(t, x, y, e)$, 
where $t$ represents timestamps, 
$x$ and $y$ denote cursor positions (relative to the top-left corner of the screen), 
and $e$ refers to mouse actions (e.g., hovers or clicks).
For more details, see the associated dataset paper~\cite{latifzadeh2025adserp}.
\section{Data representation\label{sec:data_representation}}

We cast the problem of slot attention prediction as two \gls{ml} tasks:
\begin{enumerate*}[label={(\arabic*)}]
\item \textbf{Regression analysis}: \textit{Can a user's mouse cursor trajectory predict the total fixation time on a \gls{serp} slot or even the number of fixations per slot?}
\item \textbf{Binary classification}: \textit{Can a user's mouse cursor trajectory determine whether the user noticed a particular slot?}
\end{enumerate*} 

\subsection{Cursor Data}
\label{sec:data_representation:cursor_data}

As we discuss in \Cref{sec:deep_learning_models}, 
we consider three distinct mouse cursor data embeddings: 
(1)~Transformer-based; (2)~\gls{lstm}-based; and (3)~\gls{vit}-based embeddings. 
The first two methods take as input multivariate time series, 
whereas the \gls{vit} model works with images.

\fboxsep=0pt
\fboxrule=0.5pt
\begin{figure}[!t]
  \def\w{0.28\linewidth}
  \subfloat[Heatmap]{
    \fbox{\includegraphics[width=\w]{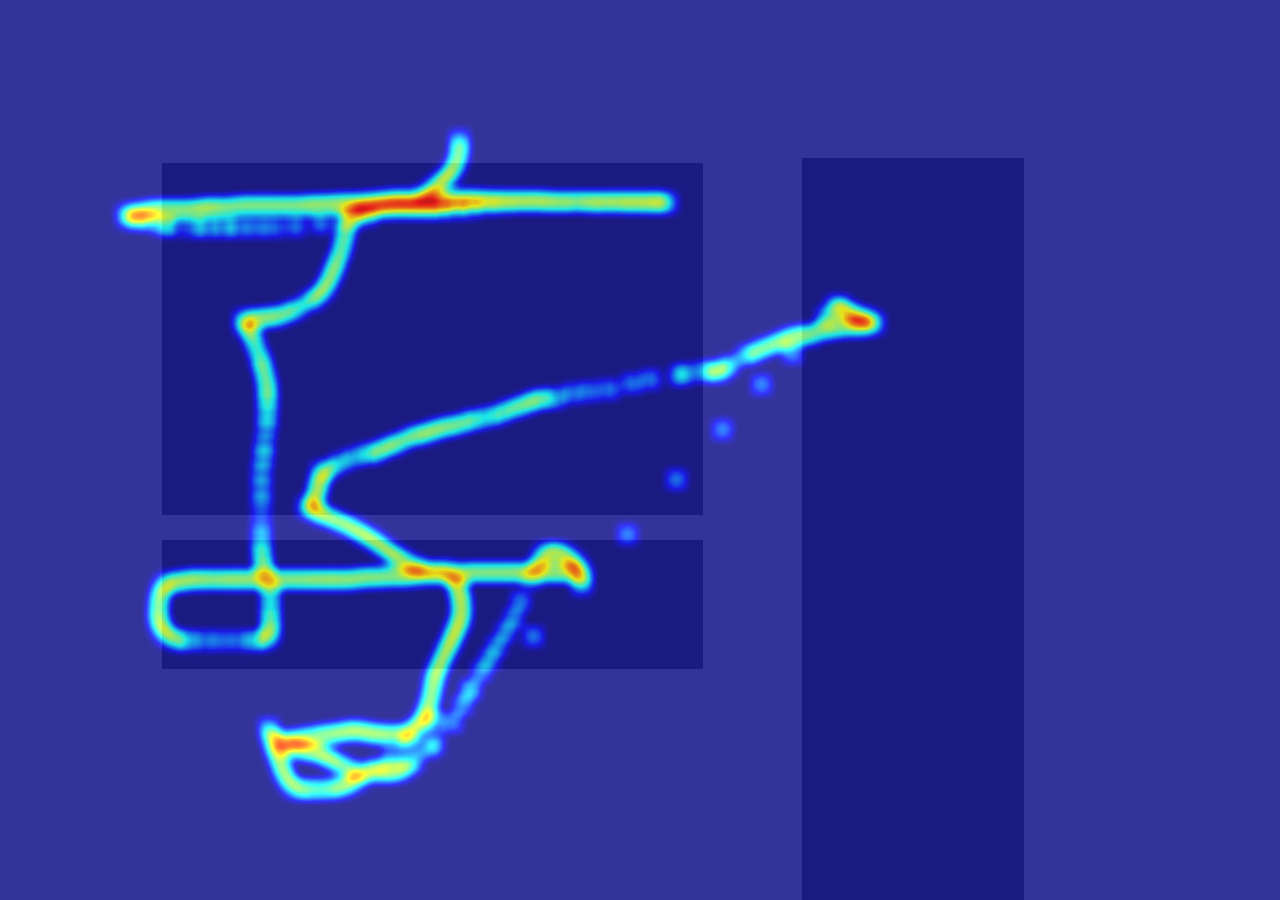}}
  }
  \hfill
  \subfloat[Colour trajectories]{
    \fbox{\includegraphics[width=\w]{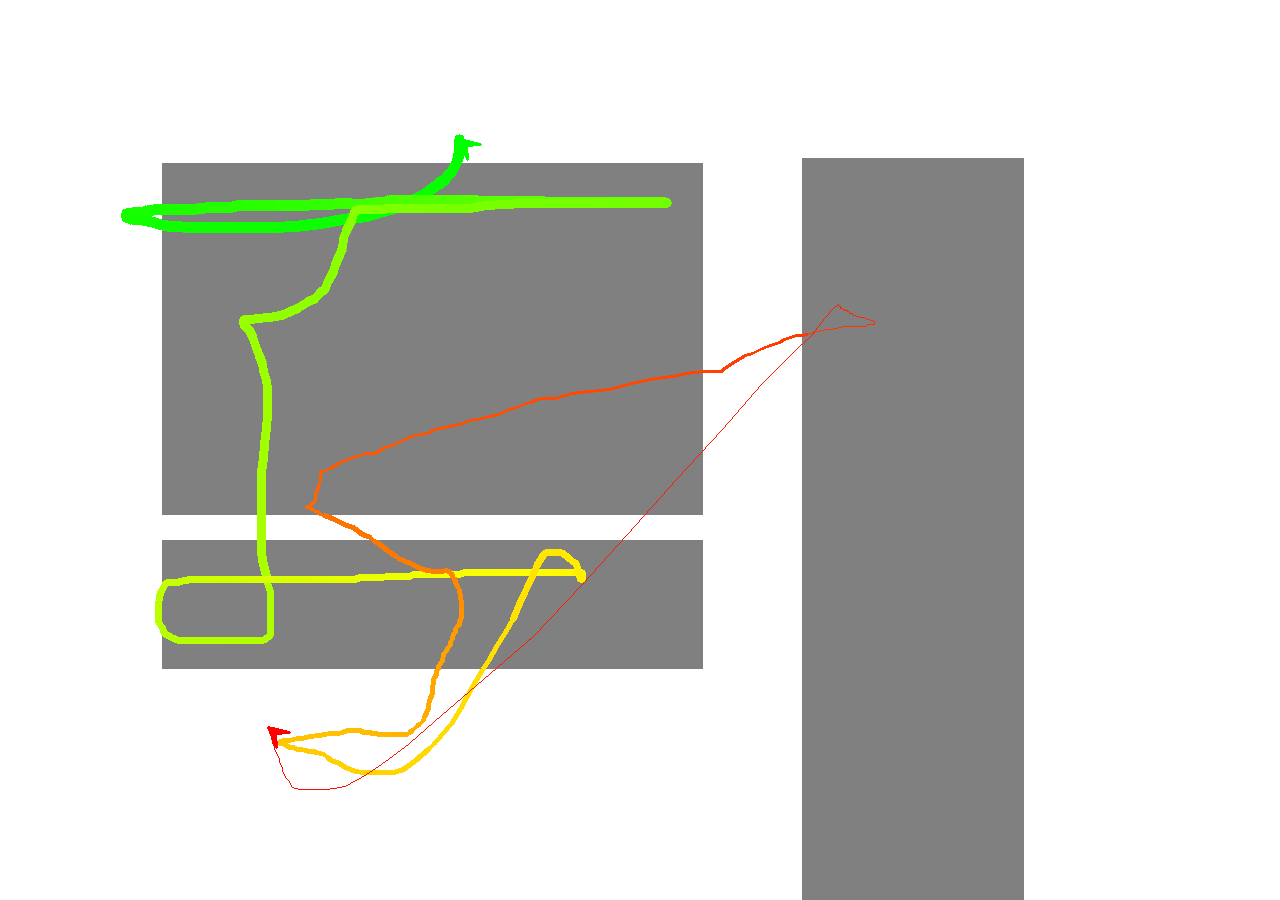}}
  }
  \hfill
  \subfloat[Colour trajectories and slot-specific colour]{
    \fbox{\includegraphics[width=\w]{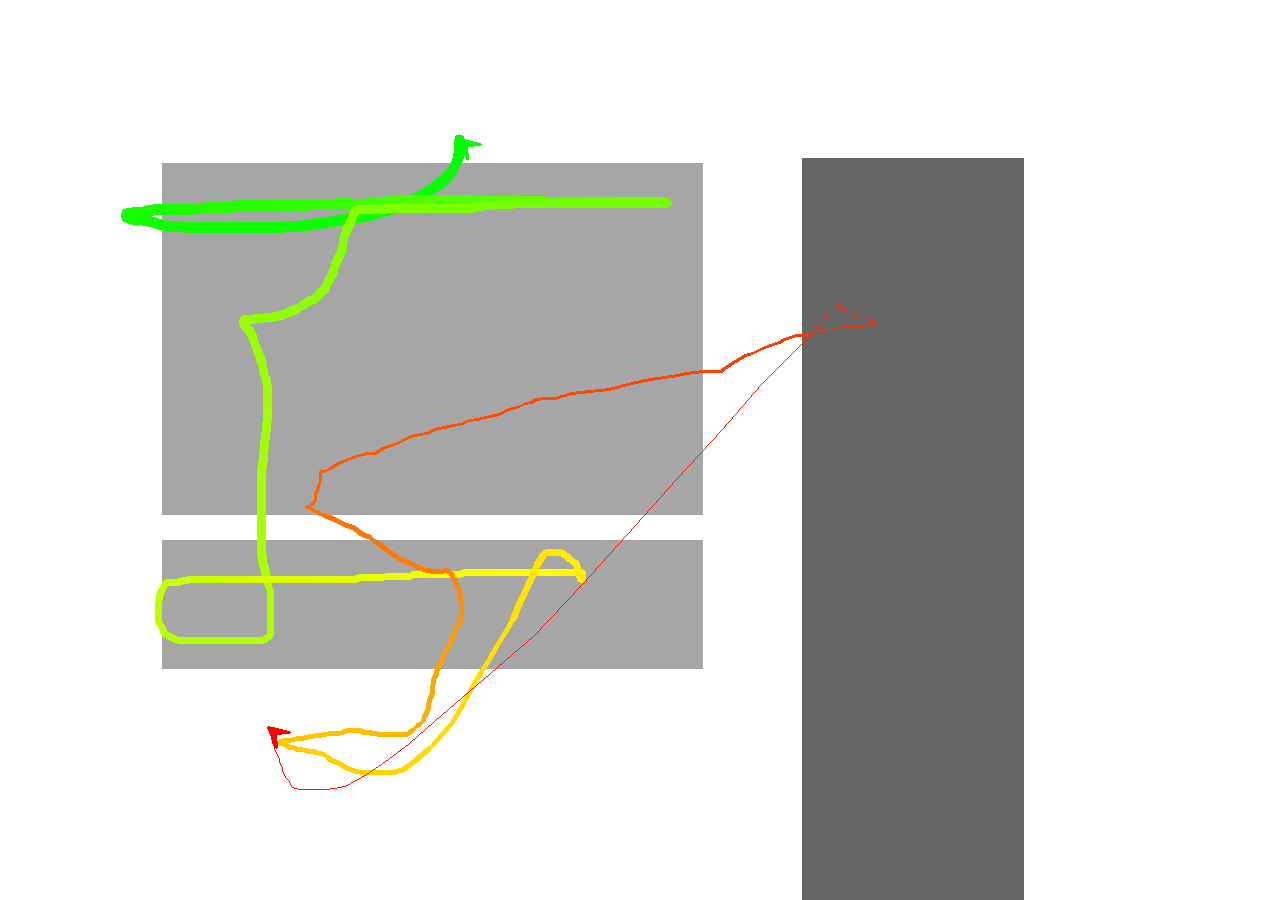}}
  }
    \caption{Visual representations of mouse movements to train the \gls{vit} models, all including slot placeholders.}
  \label{fig:representations}
\end{figure}

\subsubsection{Multivariate Time Series}
\label{sec:timeseries}

Each mouse cursor trajectory has four features at each timestep: 
cursor coordinates ($x,y$ normalized in the $[0,1]$ range w.r.t viewport size),
time spent at each coordinate, 
slot type at position,
and normalized sequence index.
The slot type is a categorical feature that takes values from $-1$ to $3$. 
A value of $-1$ indicates that the mouse cursor is positioned outside any slot, 
while values between $0$ and $3$ represent positions within specific slot categories 
(direct-top=0, direct-right=1, organic-top=2, organic-bottom=3). 
The normalized sequence index is a float in $[0,1]$
that describes the position of a sample within a sequence.\footnote{Other representations (e.g. positional encoding) yielded inferior performance.}

The data are time-ordered, with no consecutively duplicated coordinates. 
A distinguishing characteristic of this data format is its \emph{asynchronous} nature: 
unlike regular time series with constant sampling rates, mouse movement events occur irregularly. 
This happens because web browser events are first queued and then fired as soon as resources are available~\cite{maras2016secrets}.

Both Transformer and \gls{lstm} models process sequential data, but their requirements differ. 
Specifically, Transformers can handle data sequences of arbitrary and variable lengths 
so they do not require further data adjustment (e.g., padding). 
On the contrary, \glspl{lstm} require sequences of fixed length. 
We set the maximum sequence length to 250 timesteps,
a value that approximates the mean sequence length plus one standard deviation observed in our data. 
Shorter sequences are zero-padded to reach the fixed length of 250 timesteps, whereas longer sequences are truncated. 

\subsubsection{Visual Representations}

We adopt efficient representations proposed in previous work~\cite{Arapakis20_mtdl},
summarized in \Cref{fig:representations}. 
\begin{itemize}
    \item \ul{Heatmap:} Mouse coordinates are rendered using a $2D$ Gaussian kernel with a 25 pixels radius. 
    Overlapping kernels contribute cumulatively to produce aggregated heatmap values.

    \item \ul{Colour trajectories:} Mouse trajectories are rendered as lines connecting consecutive coordinates. 
    Each trajectory starts and ends with cursor-shaped markers in green and red colour, respectively. 
    Line colours follow a temperature gradient, transitioning from green at the start to red at the end. 
    Line thickness represents the percentage of time spent, 
    with thicker segments corresponding to earlier parts of the trajectory. 
    All slot boxes are shown in grey.

    \item \ul{Colour trajectories with slot-specific colours:} This representation 
    extends the previous one by using a scale of gray colours 
    to better differentiate the slot types.
\end{itemize}

\subsection{Slot Metadata}
\label{sec:data_representation:ad_metadata}

We implement the \gls{seq2seq} model architecture since
(1)~it accommodates a variable number of slots across different \glspl{serp}
and (2)~facilitates the incorporation of slot-specific features, 
enhancing both performance and generalisation capabilities. 
For the regression task, this means that the model translates a sequence of cursor movements into a sequence of predictions, 
like \gls{tft} or \gls{tfc}, for each slot. 
We note that, while the sequence of cursor movements has an inherent meaning, the order of the slots is irrelevant. 
We verified this assumption by comparing different slot ordering criteria, 
which yielded the same results (see \Cref{sec:results:regression:metadata}).

Building upon our \gls{seq2seq} model, 
we implement two features that characterise each slot: 
(1)~normalised position ($x_c$ and $y_ c$), and (2)~slot type. 
The normalised position represents the centre of the slot 
and is normalised to a range of $\left[ 0 , 1 \right]$ w.r.t viewport size. 
We explored different ways to parameterize slot positions (see \Cref{sec:results:regression:metadata_coordinates}) 
and found this representation to perform best. 
The slot type is the same categorical feature described in \Cref{sec:timeseries}.

\subsubsection{Auxiliary Slots}
\label{sec:data_representation:auxiliary_ads}

As explained in \Cref{sec:timeseries}, we defined a `slot type' feature 
that categorises the cursor's location as being either inside or outside a slot.
We noticed that the default category (-1) was not enough to detect whether 
the mouse cursor is \emph{near} organic-top or organic-bottom slots.
To address this limitation, 
we introduce \textit{auxiliary} slots that serve as other \gls{aoi}, 
enabling a more detailed categorisation of cursor positions 
and improving the overall granularity of the data representation.
In \Cref{sec:results:auxiliary_ads_impact}, we explicitly analyse the impact of auxiliary slots, 
demonstrating their effectiveness in improving model performance, 
mainly for two reasons: (1)~they refine the characterisation of cursor positions; 
and (2)~they support the learning of meaningful patterns and relationships between fixation events.

\begin{figure}[t!]
  \def\w{0.45\linewidth}
  \subfloat[]{
    \includegraphics[width=\w]{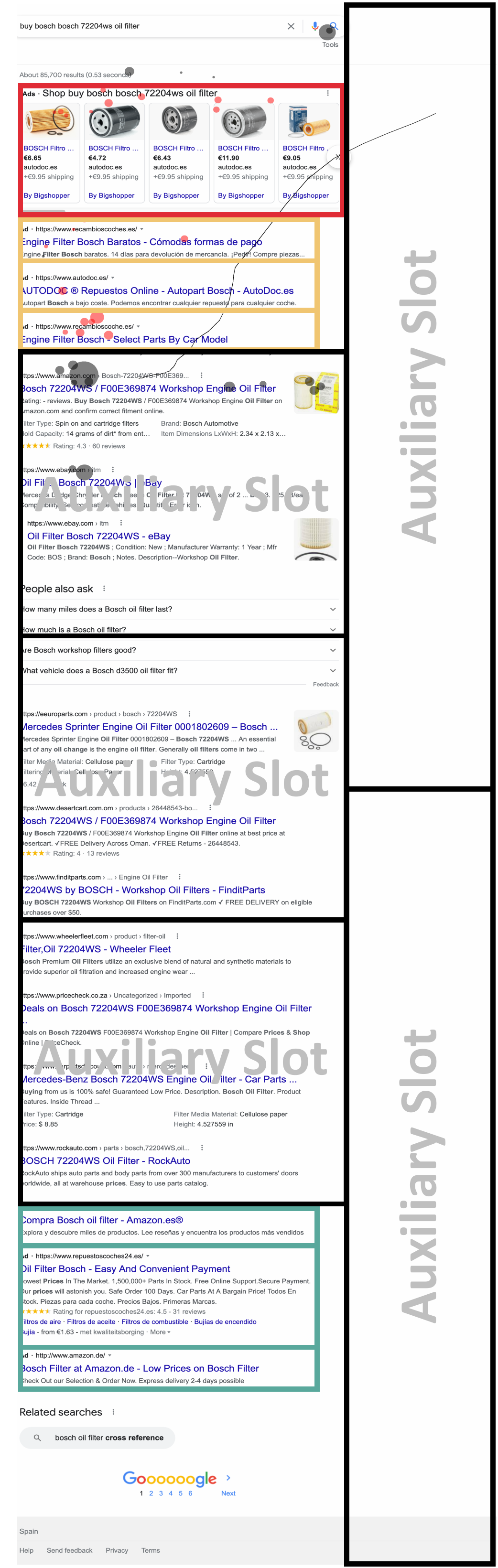}
  }
  \hfil
  \subfloat[]{
    \includegraphics[width=\w]{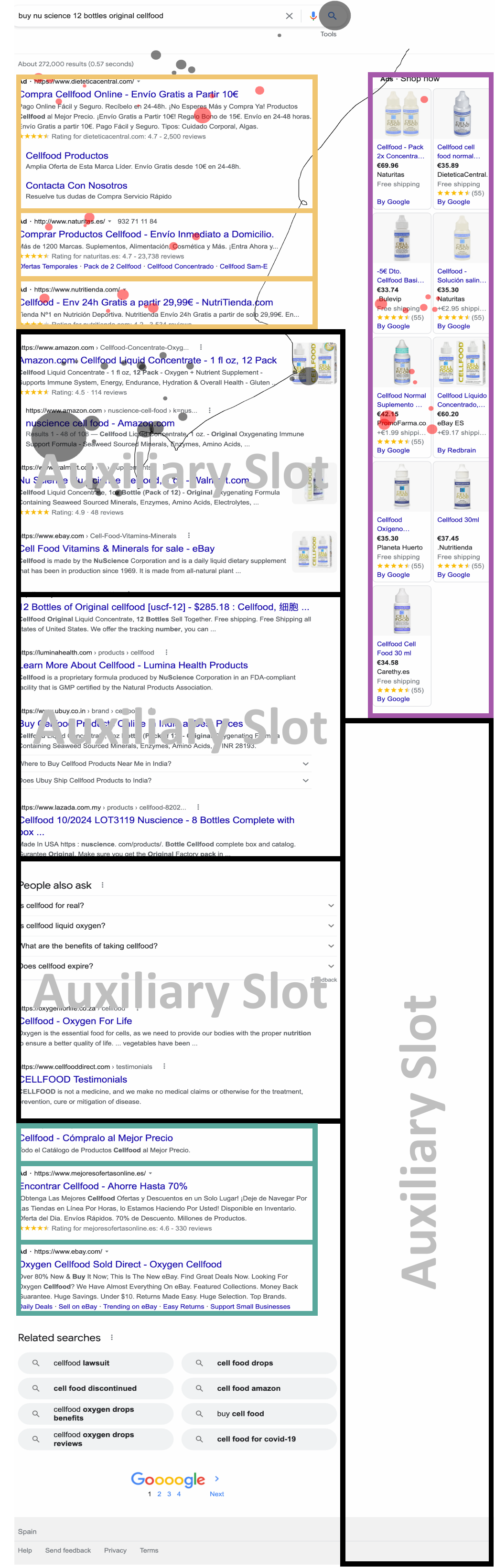}
  }
    \caption{Examples of Google \glspl{serp} with multi-slot layout. Slot categories are colour-coded ({\color{direct-top_ad_colour} direct-top slots}, {\color{direct-right_ad_colour} direct-right slots}, {\color{organic_ad_on_top_colour} organic slots on top}, and {\color{organic_ad_on_bottom_colour} organic slots on bottom}). Cursor movements are visualized as continuous black lines, while eye-tracking events are represented by filled circles (shown in \ul{black} if within auxiliary slots or \textcolor{red}{red} if within standard slots). Circle radius is proportional to event duration.}
  \label{fig:SERPs}
\end{figure}

Auxiliary slots are positioned 
as follows. Initially, the page is divided into two sections: 
the main area and the right area (see \Cref{fig:SERPs}). 
The main area is the vertical region of the web page where search results are displayed, 
along with direct-top, organic-top, and organic-bottom slots. 
The right area is the vertical portion of the web page where direct-right slots are located. 
Next, $N$ auxiliary slots are placed in the main area. 
Specifically, the space between the end of the last direct-top/organic-top slot 
and the beginning of the first organic-bottom slot 
is divided into $N$ equally-sized rectangular regions of interest, ensuring no gaps between them. 
If the page does not contain direct-top/organic-top or organic-bottom slots, 
the auxiliary slots span from the top of the web page to the bottom. 
The value of $N$ is a hyperparameter that can be optimised; 
in our setting (see \Cref{sec:results:auxiliary_ads_impact}), the optimal value is $N=3$.
Finally, for the right area we apply a simpler heuristic. 
If no direct-right slots are present, two auxiliary slots are placed: 
one in the upper-middle section and another in the lower-middle section. 
If direct-right slots are present, an auxiliary slot is positioned 
below the existing slot.

\subsection{Ground Truth}
\label{sec:data_representation:target_definition}

For both the regression and classification tasks, 
the ground truth is obtained from the eye-tracker's fixations. 
As in the case of mouse movements, fixations are time-ordered,
with no consecutively duplicated coordinates. 

\subsubsection{Regression Task}

The model is trained to predict either the \gls{tft} or the \gls{tfc} for each slot. 
Because predictions are made at the slot level, 
the number of model outputs---and thus the number of \textit{targets}---varies across \glspl{serp}. 
\gls{tft} is computed as the total duration the user spent viewing a specific slot, 
regardless of whether the fixations occurred consecutively. 
Conversely, \gls{tfc} represents the total number of fixations on a specific slot, 
independent of the duration of each fixation or their sequence. 
To ensure data quality, events lasting less than 100\,ms are excluded from both metrics, 
following recommendations from prior studies~\cite{balatsoukas2012eyetracking, liu2016predicting}.

\subsubsection{Classification Task}

The model predicts whether a user fixated on a specific category of slot. 
This involves four separate binary classification problems, one per slot category. 
Unlike in the regression task, where predictions are made at the slot level, 
the classification task produces predictions at the category level (e.g. all direct-top slots). 
The target labels are created as follows:
\begin{itemize}
    \item \ul{Cluster/Density Detection:} The longest fixation event within each slot is identified, 
    and nearby fixation events (within a reference distance) are grouped into a cluster. 
    The reference distance is dynamically determined based on the slot size. 
    This process is repeated iteratively, starting with the next longest fixation event that has not yet been assigned to a cluster. 
    By the end of this iterative process, all fixation events within a slot are associated with a cluster.
    \item \ul{Cluster/Density Characterization:} Each cluster is described using two metrics: 
    the \gls{tft} (sum of fixation durations in the cluster) 
    and the \gls{tfc} (number of fixations in the cluster).
    \item \ul{Cluster/Density Labelling:} A cluster is labelled as \texttt{True} 
    if it exceeds both \gls{tft} and \gls{tfc} thresholds, 
    which are set to the median values. 
    For categories containing multiple slots (e.g., organic-top and organic-bottom), 
    the final label is \texttt{True} if at least one slot in the category meets these criteria.
\end{itemize}
Following these steps, a user is considered to fixate 
on a slot if their gaze remains within the limited spatial area of the slot for a minimum duration, and with enough repetitions. 
This heuristic ensures that only meaningful engagement with a slot is labelled as a fixation, 
avoiding false positives due to random or brief pauses between saccadic movements. Applying this fixation labelling method, we observe fixation rates of $42\%$, $46\%$ $44\%$ and $29\%$ for the direct-top, direct-right, organic-top, and organic-bottom slots, respectively.

\section{Deep learning models}
\label{sec:deep_learning_models}

\subsection{Regression Task}
\label{sec:deep_learning_models:regression_problem}

\begin{figure*}[!ht]
  \def\w{0.44\linewidth}
  \subfloat[Baseline workflow\label{fig:baseline_model}]{
    \includegraphics[trim=0 0 0 0, clip=true, width=0.4\linewidth]{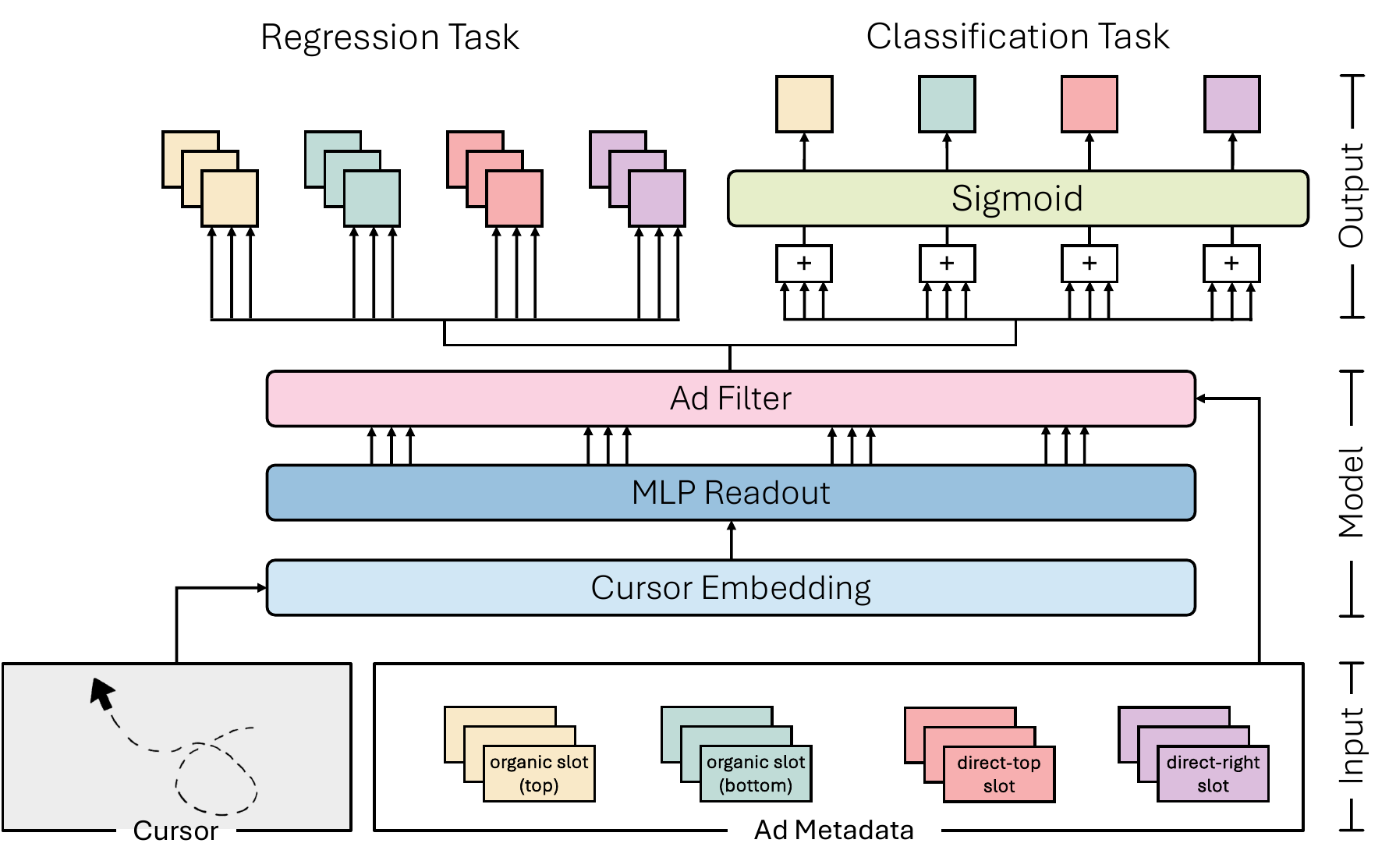}
  }
  \hspace{0.02\linewidth}
  \subfloat[\gls{seq2seq}  workflow\label{fig:seq2seq_model}]{
    \includegraphics[trim=0 0 0 0, clip=true, width=0.42\linewidth]{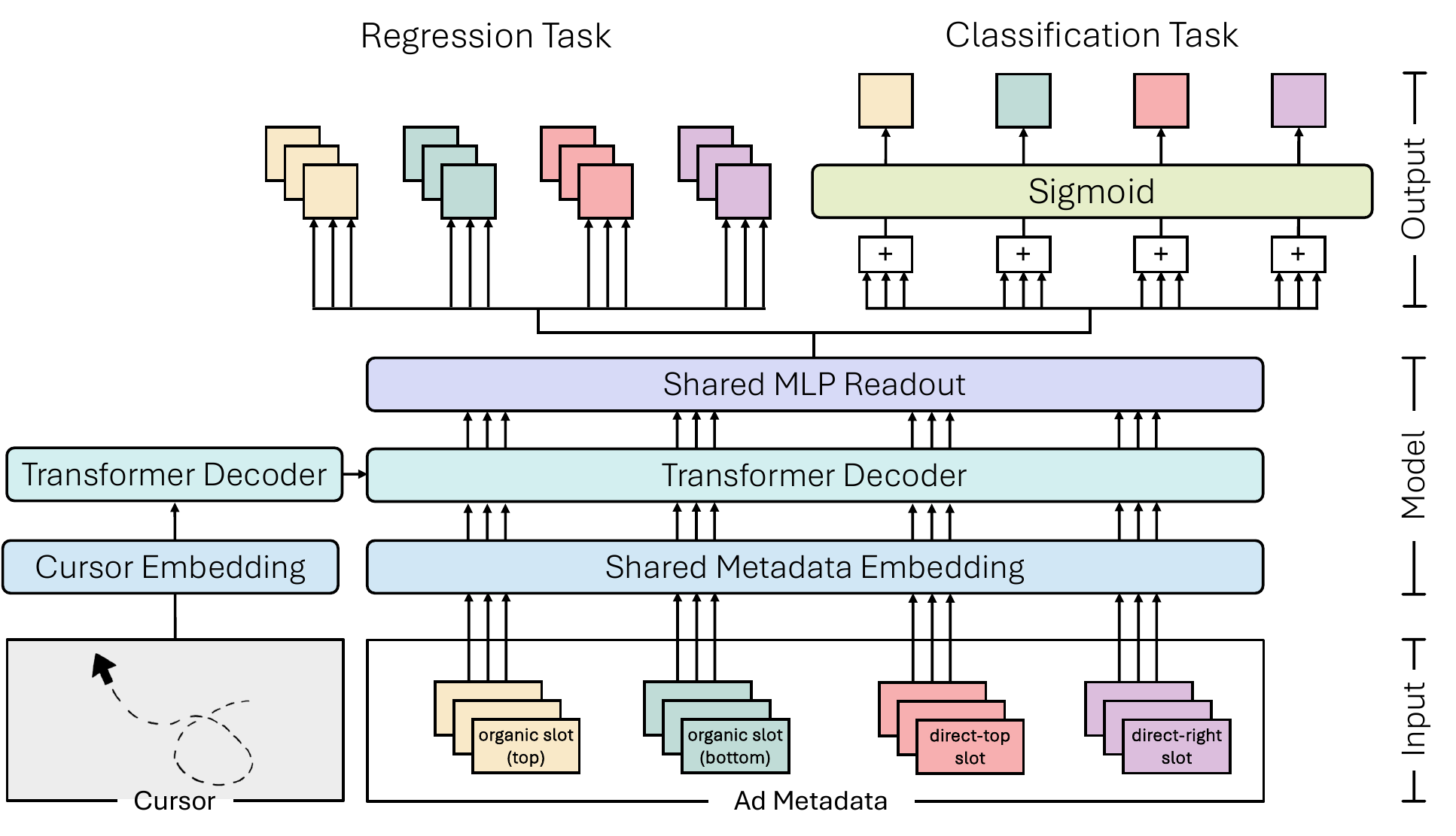}
  }
  \caption{
    Comparison of the \gls{mlp} baseline (\Cref{fig:baseline_model}) and \gls{seq2seq} approach (\Cref{fig:seq2seq_model}). The baseline embeds cursor data into a dense latent space, using an \gls{mlp} adapted to the maximum slots per document, with metadata-based filtering. The \gls{seq2seq} model projects cursor data and metadata into a shared latent space, leveraging a Transformer Encoder-Decoder for sequence relationships. Regression outputs estimate quantities (\gls{tft}/\gls{tfc}); classification aggregates scores by type and applies a sigmoid for probabilities.
  }
  \label{fig:xxx}
\end{figure*}

In this task, the model is trained to predict 
either the fixation time \gls{tft} a user spends on a specific slot or the fixation count \gls{tfc},
which refers to the number of times the user focuses on that slot. 

\subsubsection{AdSight Model}
A key challenge lies in the varying number of model outputs across trials, 
which is determined by the number of slots present in each \gls{serp}. 
To address this, we propose a \gls{seq2seq} approach 
that maps a sequence of cursor movements to a corresponding sequence of predicted fixation times. 
We use an Encoder-Decoder Transformer, which is capable of modelling the relationships between sequences of arbitrary length. 
Additionally, we incorporates slot-specific features or metadata, 
which significantly improves the model's performance (see \Cref{sec:results}). 

As illustrated in \Cref{fig:seq2seq_model}, the process begins with the model accepting two primary inputs: 
(1)~the mouse cursor movement data; and (2)~the slot-specific metadata. 
These inputs are projected into a shared latent space using different 
embeddings, one for cursor data and another for metadata. 
Next, the cursor and metadata representations are fed into an Encoder-Decoder Transformer 
designed to capture both intra- and inter-sequence relationships. 
The encoder processes the sequence of cursor embeddings, 
while the decoder takes as input the sequence of slot metadata embeddings and 
the encoder's output. 
Finally, the decoder's output is passed through a shared \gls{mlp}, applied 
to each slot in the sequence. 
The term ``shared \gls{mlp}'' in \Cref{fig:seq2seq_model} 
denotes that a single instance of the \gls{mlp}, with identical weights, 
is used to predict the \gls{tft}/\gls{tfc} for all slots. 

\noindent{\textbf{Cursor Embedding.}} 
For \ul{time series data}, the embedding process involves two steps. 
First, a shared \gls{mlp} independently projects the features of each mouse cursor position into a common latent space. 
This \gls{mlp} consists of one hidden and an output layer, 
both with a number of neurons equal to the latent space size, $l \in \left\lbrace 16, 32, 64 \right\rbrace$. 
This initial step focuses on capturing the relationships among the features of individual cursor positions.
The second step improves the latent representations of mouse cursor positions by incorporating intra-sequence relationships. 
We consider two different architectures for this step: 
(1) a bidirectional \gls{lstm} network; and (2) a Transformer encoder. 
For both architectures, the implementation consists of two layers, 
with the hidden space dimension set equal to the latent space size. 
For the Transformer encoder, we use two attention heads and a feed-forward network size of $512$ 
(refer to \cite{vaswani2017attention} for details). 
The final output of the mouse cursor embedding is a sequence with a length equal to the number of cursor movements, 
where each element in the sequence is represented by $l$ features. 

For \ul{image data}, the embedding process implements the \gls{vit} model 
in \cite{dosovitskiy2021an}. 
The entire architecture is frozen except for the last layer, 
which is replaced by two trainable dense layers with $k \cdot l$ and $l$ neurons, respectively, 
where $k \in \left\lbrace 2 , 4 , 8 \right\rbrace$. 
The final output of the cursor embedding can be seen as a sequence of only one element, 
the summary of the cursor trajectory, characterized by $l$ features. 

\noindent{\textbf{Slot Metadata Embedding.}} This embedding relies on a shared neural network 
that projects each slot into a common latent space. 
Initially, each slot feature is independently projected into an intermediate latent space of size $l$. 
For continuous features ($x_c$ and $y_c$), we use a linear layer followed by a \gls{relu} activation function, 
whereas for categorical features (e.g., slot type) we use a word embedding layer.\footnote{A simple lookup table 
that stores embeddings of a fixed dictionary and size.}
Next, the resulting embeddings are concatenated into a $3 \cdot l$ tensor 
and further projected into the final latent space of size $l$ 
using a dense layer with \gls{relu} activation. 
The output of the metadata embedding is a sequence with a length equal to the number of slots on the \gls{serp}
(including auxiliary slots), where each element in the sequence is represented by $l$ features.

\noindent{\textbf{Encoder-Decoder Transformer.}} This component 
captures the intra- and inter-sequence relationships between the two incoming data streams. 
Specifically, the encoder processes the sequence of cursor embeddings, 
while the decoder takes as input both the sequence of slot metadata embeddings and the output of the encoder. 
The decoder outputs an enhanced representation of the sequence of slots.
We use two encoder-decoder layers with a latent space equal to $l$, 
two attention heads and a feedforward network size of $512$. 

\noindent{\textbf{Shared \gls{mlp} Readout.}} It consists of a hidden dense layer followed by an output dense layer, 
with $l$ and $1$ neuron(s), respectively. 
Both layers utilize \gls{relu} activation functions. 
The use of a \gls{relu} activation function in the output layer 
is motivated by the need to predict fixation time and fixation counts, 
which are both non-negative quantities. 
Empirical tests showed that removing the \gls{relu} activation in the output layer degrades model performance. 
Moreover, this \gls{mlp} is shared across all elements in the slot sequence. 
This approach reduces the number of trainable parameters, prevents overfitting, and improves the model's generalization capabilities. 

\noindent{\textbf{Loss Function.}} 
We explore two distinct loss functions: 
(1) the \gls{mse}; and (2) the Listwise Rank Loss \cite{Cao2007Learning}. 
While \gls{mse} evaluates the error in predicting the exact fixation time/count for each slot on the page,
the Listwise Rank Loss assesses how well the model's predictions replicate the slot ranking based on true fixation time/count. 
Depending on the specific use case, one of these loss functions may be more suitable than the other.

For \gls{mse}, the total loss is computed as the sum of the \gls{mse} values across all slot categories, 
which may also include auxiliary slots. 
We introduce a hyperparameter $0 \leq \alpha \leq 1$ 
to control the contribution of the auxiliary slots to the overall loss function. 
When $\alpha=0$, the error from auxiliary slot predictions does not influence the model update, 
whereas $\alpha=1$ means the error from auxiliary slots 
is weighted equally with the four slot categories. 
As shown in \Cref{sec:results:auxiliary_ads_impact}, incorporating $\alpha>0$ provides the model with certain advantages. 
With respect to Listwise Rank Loss, the total loss is computed 
by adding the Listwise Rank Loss for the slot categories 
and the Listwise Rank Loss for the auxiliary slots. 
The contribution of the auxiliary slots is weighted by the hyperparameter $\alpha$. 
The loss functions are optimized using Adam optimizer (stochastic gradient descent with momentum) 
with learning rate $\eta \in \left\lbrace 10^{-3}, 10^{-4}, 10^{-5}\right\rbrace$, 
and decay rates $\beta_{1}=0.9$ and $\beta_{2}=0.999$. 
We set a maximum number of $100$ epochs, 
using an early stopping of $25$ epochs that monitors the validation loss, 
and tried different batch sizes $b \in \left\lbrace 16 , 32 , 64 \right\rbrace$. 

\noindent{\textbf{Hyperparameter Optimisation.}} The total hyperparameter space consists of the latent space size ($l$), the hidden layer factor ($k$, which applies only to the \gls{vit} cursor embedding), the weight of the auxiliary slot loss ($\alpha$), the learning rate ($\eta$), and the batch size ($b$). We optimize these hyperparameters using Bayesian optimization through the Optuna library \cite{akiba2019optuna}, by implementing 3-fold cross-validation.

\subsection{Classification Task}
\label{sec:deep_learning_models:binary_problem}

In this task, we leverage the mouse cursor trajectories to determine whether the user has noticed a specific category of slot. 
Specifically, we address four binary classification problems, 
corresponding to the four slot categories under consideration: 
direct-top, direct-right, organic-top, and organic-bottom. 
We propose a unified model capable of producing all four binary scores simultaneously.

The proposed approach builds on the workflow described in the regression case, 
with an additional post-processing step applied to the outputs (see \Cref{fig:seq2seq_model}). 
As previously discussed
, the workflow generates as many outputs as number of slots in a given trial. 
For the binary classification use case, we aggregate the outputs corresponding to the same slot category using metadata. 
This aggregation yields a score for each slot category including the auxiliary slots. 
Finally, a sigmoid activation function is applied to each score, 
resulting in a probability $p$ for the user’s attention to a specific slot category. 
A prediction of  $p > 0.5$ indicates that the user noticed the corresponding slot category. 
This approach is 
beneficial due to the uneven distribution of slots within our dataset. 
For instance, since only 31\% of trials contain a direct-right slot, training four separate binary models would result in the direct-right slot model being trained on just 31\% of the available data. This could lead to potential overfitting and poor generalisation. Moreover, creating individual models adds 
computational cost 
and diminishes maintainability.

\noindent{\textbf{Loss Function.}} The total loss is calculated as the sum of five binary crossentropy, 
one for each slot category including the auxiliary slots. 
As in the regression case, the contribution of the auxiliary loss is weighted with a hyperparameter $0 \leq \alpha \leq 1$.\footnote{We also attempted to adjust the weights of the direct-top, direct-right, organic-top, and organic-bottom slot losses, considering the distribution of these categories in the dataset. However, no statistically significant improvement was observed.} 
Optimization is performed using the Adam optimizer with a learning rate $\eta \in \left\lbrace 10^{-3}, 10^{-4}, 10^{-5}\right\rbrace$, 
and decay rates $\beta_{1}=0.9$ and $\beta_{2}=0.999$. 
We set a maximum number of $100$ epochs, using an early stopping of $25$ epochs that monitors the validation loss, 
and tested different batch sizes $b \in \left\lbrace 16 , 32 , 64 \right\rbrace$. 

\noindent{\textbf{Hyperparameter Optimisation.}} The total hyperparameter space is similar with the regression use case: 
latent space size ($l$), the hidden layer factor ($k$, which applies only to the \gls{vit} cursor embedding), the weight of the auxiliary slot loss ($\alpha$), the learning rate ($\eta$), and the batch size ($b$). Also in this case, we optimize these hyperparameters using Bayesian optimization through the Optuna library \cite{akiba2019optuna}, by implementing 3-fold cross-validation.

\subsection{Baselines}
\label{sec:baselines}

We perform a comparison with baseline models based on the workflow shown in \Cref{fig:baseline_model}. 
The main 
differences between the \gls{seq2seq} workflows and the baselines are: 
(1)~replacement of the Encoder-Decoder transformer layer with an \gls{mlp}; and (2)~the use of slot metadata solely for filtering the model's output. 
The remainder of the workflow---cursor embedding, binary post-processing, loss functions, optimizer, and hyperparameter optimization---is identical for the \gls{seq2seq} framework.

As indicated in \Cref{fig:baseline_model}, mouse cursor movement data are projected into a latent space 
using the same embeddings employed in the \gls{seq2seq} case: \gls{lstm}, Transformer encoder, or \gls{vit}. 
The cursor embedding output is then flattened into a tensor of size $L \cdot l$ for LSTM and the Transformer encoder, 
or $l$ for \gls{vit}, where $l$ denotes the latent space size and $L$ represents the cursor movement sequence length. 
This tensor is then fed into an \gls{mlp} with $h \in \left[ 2, 3, 4\right]$ dense layers, 
each followed by a \gls{relu} activation function. 
The first layer contains $L \cdot l$ or $l$ neurons, depending on the embedding used, 
while the hidden layers have $l$ neurons. 
The output layer has $14$ neurons, due to the nature of the baseline workflow. 
Unlike standard dense layers that produce a fixed number of outputs, 
our experimental datasets vary in the number of model predictions depending on the number of slots in each trial. 
To address this variability, the \gls{mlp}'s output layer 
is configured to match the maximum number of slots ($14$) in a single trial. 
Consequently, the baseline workflow’s readout \gls{mlp} always generates $14$ outputs, 
and slot metadata are subsequently applied to filter predictions corresponding to the actual slots in each document. 
For the classification task, the same postprocessing steps described in \Cref{sec:deep_learning_models:binary_problem} 
are applied to the filtered predictions, yielding the final probability $p$ for the user’s attention to a specific slot category.

The classification problem has been explored in prior research~\cite{Arapakis20_mtdl}, prompting us to compare and contrast existing models. BLSTM demonstrated superior average performance for different slot types when cursor data were represented as time series. Conversely, ResNet50 performed best when cursor data were encoded as \textit{Heatmaps} or \textit{Colour trajectories}. Building on these insights, we include both BLSTM and ResNet50 in our experiments, using the configurations reported by \citet{Arapakis20_mtdl}. We note that, while these models incorporate cursor embeddings similar to our proposed method, they do not utilise \gls{seq2seq} layers. Additionally, \citet{Arapakis20_mtdl} collected self-reported ground truth labels.

\subsection{Performance Metrics}
\label{sec:metrics}

For the regression task, we evaluate model performance using \gls{mse} and \gls{ndcg}. \gls{mse} quantifies prediction error by averaging the squared differences between predicted and actual values, while \gls{ndcg} measures how well the predicted rankings align with the ground truth. For the classification task, we use \gls{auc} and the F1 score. \gls{auc} captures the model's ability to distinguish between positive and negative classes by integrating true and false positive rates across thresholds. The F1 score, the harmonic mean of precision and recall, balances the correctness and completeness of positive predictions. Results represent the average predictive and ranking performance across test instances.
\section{Results\label{sec:results}}

\begin{table*}[!ht]
\caption{
    Regression results. 
    The first and second best results are highlighted 
    in bold and underline format.
}
\label{tab:time}
\resizebox{.935\linewidth}{!}{
\begin{tabular}{ll|cc|cc|cc|cc}
    \toprule
& & \multicolumn{4}{c|}{\gls{tft}} & \multicolumn{4}{c}{\gls{tfc}} \\
\cmidrule(lr){3-6} \cmidrule(lr){7-10}
Readout & Embedding & \gls{mse} $\downarrow$ & NDCG $\uparrow$ & Rank Loss $\downarrow$ & NDCG $\uparrow$ & \gls{mse} $\downarrow$ & NDCG $\uparrow$ & Rank Loss $\downarrow$ & NDCG $\uparrow$ \\
    \midrule
\multirow{5}{*}{\gls{mlp}} & Transformer & $4.99 \pm 0.03$ & $82.36 \pm 0.03$ & $0.2482 \pm 0.0005$ & $89.51 \pm 0.02$& $79.50 \pm 0.04$ & $83.80 \pm 0.04$ & $0.2113 \pm 0.0007$ & $89.91 \pm 0.03$ \\
& \gls{lstm}                            & $5.04 \pm 0.03$ & $81.70 \pm 0.04$ & $0.2499 \pm 0.0004$ & $89.46 \pm 0.02$& $84.10 \pm 0.05$ & $82.66 \pm 0.03$ & $0.2141 \pm 0.0005$ & $89.83 \pm 0.02$ \\
& \gls{vit}                             & $5.16 \pm 0.04$ & $80.74 \pm 0.03$ & $0.2551 \pm 0.0003$ & $86.87 \pm 0.05$& $84.63 \pm 0.03$ & $81.45 \pm 0.02$ & $0.2211 \pm 0.0002$ & $87.45 \pm 0.02$ \\
& \gls{vit}-colour                      & $5.11 \pm 0.03$ & $80.88 \pm 0.03$ & $0.2537 \pm 0.0005$ & $86.92 \pm 0.04$& $84.54 \pm 0.02$ & $81.50 \pm 0.02$ & $0.2207 \pm 0.0004$ & $87.47 \pm 0.02$ \\
& \gls{vit}-heatmap                     & $5.15 \pm 0.04$ & $80.75 \pm 0.02$ & $0.2505 \pm 0.0003$ & $87.10 \pm 0.06$& $84.69 \pm 0.03$ & $81.12 \pm 0.04$ & $0.2152 \pm 0.0003$ & $87.75 \pm 0.04$ \\ \midrule
\multirow{5}{*}{\gls{seq2seq}} & Transformer   & $\bm{2.86 \pm 0.02}$ & $\bm{96.07 \pm 0.04}$ & $\bm{0.1881 \pm 0.0004}$ & $\bm{96.08 \pm 0.04}$& $\bm{50.07 \pm 0.04}$ & $\bm{96.36 \pm 0.05}$ & $\bm{0.1176 \pm 0.0004}$ & $\bm{96.45 \pm 0.05}$ \\
& \gls{lstm}                            & \underline{$3.19 \pm 0.03$} & \underline{$95.84 \pm 0.02$} & $0.1902 \pm 0.0002$ & $95.85 \pm 0.02$& $51.76 \pm 0.03$ & $95.88 \pm 0.03$ & $0.1181 \pm 0.0003$ & $96.03 \pm 0.02$ \\
& \gls{vit}                             & $3.34 \pm 0.02$ & $95.64 \pm 0.03$ & $0.1903 \pm 0.0003$ & $95.82 \pm 0.03$& $51.17 \pm 0.02$ & $95.90 \pm 0.02$ & $0.1191 \pm 0.0003$ & $95.93 \pm 0.03$ \\
& \gls{vit}-colour                      & $3.30 \pm 0.03$ & $95.68 \pm 0.02$ & $0.1901 \pm 0.0003$ & $95.87 \pm 0.02$& \underline{$50.98 \pm 0.03$} & \underline{$96.04 \pm 0.02$} & \underline{$0.1180 \pm 0.0004$} & \underline{$96.08 \pm 0.03$} \\
& \gls{vit}-heatmap                     & $3.31 \pm 0.04$ & $95.65 \pm 0.03$ & \underline{$0.1900 \pm 0.0003$} & \underline{$95.88 \pm 0.02$}& $51.09 \pm 0.02$ & $95.96 \pm 0.02$ & $0.1185 \pm 0.0002$ & $95.98 \pm 0.02$ \\
\bottomrule
\end{tabular}}
\end{table*}

\begin{table*}[!ht]
\caption{
    Classification results. 
    The first and second best results are highlighted 
    in bold and underline format.
}
\label{tab:binary}
\resizebox{.935\textwidth}{!}{
\begin{tabular}{ll|cccccccc|cc}
    \toprule
    & & \multicolumn{2}{c}{Direct-Top} & \multicolumn{2}{c}{Direct-Right} & \multicolumn{2}{c}{Organic-Top} & \multicolumn{2}{c}{Organic-Bottom} & \multicolumn{2}{|c}{Average Performance} \\ \cmidrule(lr){3-4} \cmidrule(lr){5-6} \cmidrule(lr){7-8} \cmidrule(lr){9-10} \cmidrule(lr){11-12}
    Readout & Embedding
    & \gls{auc} $\uparrow$ & F1 $\uparrow$ & \gls{auc} $\uparrow$ & F1 $\uparrow$ & \gls{auc} $\uparrow$ & F1 $\uparrow$ & \gls{auc} $\uparrow$ & F1 $\uparrow$ & \gls{auc} $\uparrow$ & F1 $\uparrow$ \\
    \midrule
\multirow{5}{*}{\gls{mlp}} & Transformer   & $71.59 \pm 0.05$ & $64.40 \pm 0.04$ & $72.22 \pm 0.05$ & $65.27 \pm 0.06$ & $62.57 \pm 0.05$ & $57.67 \pm 0.06$ & $76.58 \pm 0.06$ & $73.00 \pm 0.04$ & $71.95 \pm 0.06$ & $66.67 \pm 0.04$  \\
 &  \gls{lstm}                               & $69.65 \pm 0.04$ & $62.71 \pm 0.07$ & $68.02 \pm 0.06$ & $61.17 \pm 0.04$ & $59.53 \pm 0.06$ & $55.30 \pm 0.03$ & $74.14 \pm 0.05$ & $70.40 \pm 0.06$ & $69.30 \pm 0.04$ & $64.16 \pm 0.06$  \\
 &  \gls{vit}                                & $64.78 \pm 0.07$ & $58.18 \pm 0.05$ & $63.75 \pm 0.06$ & $59.16 \pm 0.07$ & $59.97 \pm 0.06$ & $55.92 \pm 0.07$ & $78.28 \pm 0.06$ & $73.08 \pm 0.05$ & $69.11 \pm 0.04$ & $64.10 \pm 0.05$  \\
 &  \gls{vit}-colour                         & $65.04 \pm 0.04$ & $59.97 \pm 0.06$ & $64.25 \pm 0.07$ & $59.66 \pm 0.06$ & $61.36 \pm 0.05$ & $56.47 \pm 0.04$ & $78.31 \pm 0.04$ & $73.18 \pm 0.07$ & $69.52 \pm 0.05$ & $64.31 \pm 0.07$  \\
 &  \gls{vit}-heatmap                        & $64.97 \pm 0.05$ & $59.20 \pm 0.07$ & $64.26 \pm 0.06$ & $59.80 \pm 0.05$ &  $58.65 \pm 0.07$ & $54.68 \pm 0.05$ & $78.21 \pm 0.07$ & $73.03 \pm 0.06$ & $68.95 \pm 0.06$ & $63.86 \pm 0.06$  \\

\midrule
\multirow{5}{*}{\gls{seq2seq}} & Transformer       & $\bm{80.79 \pm 0.05}$ & $\bm{73.90 \pm 0.07}$ & $\bm{81.72 \pm 0.06}$ & $\bm{75.07 \pm 0.05}$ & $\bm{71.87 \pm 0.04}$ & $\bm{67.27 \pm 0.06}$ & $85.85 \pm 0.04$ & $82.57 \pm 0.04$ & $\bm{81.24 \pm 0.06}$ & $\bm{76.25 \pm 0.05}$  \\
 &  \gls{lstm}        & \underline{$79.05 \pm 0.06$} & \underline{$72.41 \pm 0.04$} & \underline{$77.52 \pm 0.07$} & \underline{$70.97 \pm 0.07$} & $69.09 \pm 0.03$ & $65.17 \pm 0.03$ & $83.61 \pm 0.06$ & $80.17 \pm 0.07$ & $78.77 \pm 0.05$ & $73.93 \pm 0.03$  \\
 &  \gls{vit}         & $74.58 \pm 0.02$ & $68.28 \pm 0.05$ & $73.65 \pm 0.03$ & $69.36 \pm 0.04$ & $69.60 \pm 0.03$ & $65.86 \pm 0.04$ & \underline{$87.85 \pm 0.03$} & \underline{$82.90 \pm 0.05$} & $78.79 \pm 0.03$ & $74.09 \pm 0.03$  \\
 &  \gls{vit}-colour  & $74.64 \pm 0.03$ & $69.10 \pm 0.04$ & $73.85 \pm 0.04$ & $69.56 \pm 0.03$ & \underline{$71.03 \pm 0.06$} & \underline{$66.44 \pm 0.05$} & $\bm{87.95 \pm 0.02}$ & $\bm{83.11 \pm 0.04}$ & \underline{$79.15 \pm 0.04$} & \underline{$74.24 \pm 0.04$}  \\
 &  \gls{vit}-heatmap & $74.67 \pm 0.02$ & $69.97 \pm 0.03$ & $73.69 \pm 0.04$ & $69.53 \pm 0.04$ & $68.31 \pm 0.07$ & $64.64 \pm 0.04$ & $\bm{87.95 \pm 0.03}$ & $\bm{83.11 \pm 0.02}$ & $78.62 \pm 0.04$ & $73.83 \pm 0.02$  \\

\midrule
\multicolumn{2}{c|}{BiLSTM~\cite{Arapakis20_mtdl}} & $63.27 \pm 0.06$ & $62.04 \pm 0.04$ & $65.02 \pm 0.05$ & $60.73 \pm 0.04$ & $57.46 \pm 0.04$ & $54.72 \pm 0.04$ & $72.81 \pm 0.05$ & $69.86 \pm 0.05$ & $66.20 \pm 0.03$ & $63.29 \pm 0.06$ \\
\multicolumn{2}{c|}{ResNet50~\cite{Arapakis20_mtdl}}          & $63.76 \pm 0.03$ & $57.79 \pm 0.03$ & $63.51 \pm 0.06$ & $58.33 \pm 0.05$ & $60.78 \pm 0.05$ & $55.30 \pm 0.06$ & $76.64 \pm 0.04$ & $71.33 \pm 0.06$ & $68.29 \pm 0.06$ & $62.48 \pm 0.05$ \\
\multicolumn{2}{c|}{ResNet50-colour~\cite{Arapakis20_mtdl}}   & $64.27 \pm 0.05$ & $58.11 \pm 0.03$ & $63.82 \pm 0.06$ & $58.83 \pm 0.04$ & $60.99 \pm 0.05$ & $55.53 \pm 0.04$ & $76.99 \pm 0.03$ & $71.61 \pm 0.04$ & $68.65 \pm 0.05$ & $62.79 \pm 0.06$ \\
\multicolumn{2}{c|}{ResNet50-heatmap~\cite{Arapakis20_mtdl}}  & $65.40 \pm 0.04$ & $59.42 \pm 0.05$ & $60.89 \pm 0.04$ & $59.24 \pm 0.04$ & $59.40 \pm 0.07$ & $53.78 \pm 0.05$ & $77.06 \pm 0.04$ & $72.34 \pm 0.06$ & $68.29 \pm 0.03$ & $63.17 \pm 0.05$ \\

\bottomrule
\end{tabular}}
\end{table*}

\subsection{Regression Task\label{sec:results:regression}}

\Cref{tab:time} compares the performance of AdSight 
(\gls{seq2seq} readout) against the baselines (\gls{mlp} readout) in predicting \gls{tft} and \gls{tfc} 
across different cursor data representations and embeddings. 
Columns 3-4 and 7-8 present results optimised using the \gls{mse} loss function,
while columns 5-6 and 9-10 display results optimized with the rank loss function. 
In the \textit{Embedding} column, ViT, ViT-colour, and ViT-heatmap 
refer to the ViT embedding applied to cursor data represented as: 
colour trajectories, colour trajectories and slot-specific colour, and heatmap representations, respectively. 

We observe that the \gls{seq2seq} approach consistently outperforms the \gls{mlp} baseline, 
regardless of the target (\gls{tft} or \gls{tfc}), loss function (\gls{mse} or rank loss), 
and cursor data representation and embedding. 
Our findings further suggest that using a time series cursor data representation 
with a Transformer encoder-based cursor embedding yields the best results, 
irrespective of the target metric, read-out selection (\gls{seq2seq} or \gls{mlp}), or loss function.
We further observe that the \gls{seq2seq} model, 
utilising time series cursor data representation and Transformer encoder embeddings, 
is the best performing model regardless of the target metric or loss function. 
For example, for \gls{tft} prediction, the \gls{seq2seq} model achieves an average \gls{mse} of $2.86$, 
corresponding to an average error of $1.69$ seconds in predicting user attention on a specific slot. 
Moreover, all the visual representations of cursor data yield similar results, 
with ViT-colour being the approach that tends to achieve slightly better outcomes.  

Considering the \gls{ndcg} results, we note some interesting patterns. 
Specifically, regardless of the target, a model optimised using the rank loss function 
tends to achieve a better \gls{ndcg} than a model optimized using \gls{mse}. 
Also, irrespective of the loss function, a model predicting \gls{tfc} 
generally obtains better \gls{ndcg} than a model predicting \gls{tft}. 
The Wilcoxon signed-rank test (with correction for multiple testing) revealed that
models with \gls{seq2seq} readout outperformed those with \gls{mlp} readout
across all target/metric/embedding combinations ($p < .05$). 
Additionally, for each target/metric, 
the combination of \gls{seq2seq} readout and Transformer embedding
was always statistically superior to the second-best model ($p < .05$).

In what follows, we examine the influence of slot metadata, cursor features, and auxiliary slots on model performance. For simplicity and brevity, we focus on the best model (\gls{seq2seq} readout, time series cursor data representation, and Transformer cursor embedding) predicting \gls{tft} and \gls{mse}. Similar results can be observed for \gls{tfc} and/or when optimizing the rank loss function.

\subsubsection{Slot coordinate impact}
\label{sec:results:regression:metadata_coordinates}

\Cref{tab:metadata:coordinate} summarizes the impact 
of the slot coordinate parametrization 
(keeping the rest of the parameters constant) 
on \gls{mse} and \gls{ndcg},
considering three cases: 
(1)~normalized center, 
(2)~extreme values ($x_{min}$, $x_{max}$, $y_{min}$, $y_{max}$) 
and (3)~$x_{min}$, $x_{max}$ together with  width and height. 
As shown in the table, the normalised center coordinates yield the best results. 

\subsubsection{Slot metadata impact}
\label{sec:results:regression:metadata}

\begin{table}[b!]
\begin{minipage}[t]{0.51\linewidth}
\vspace*{0pt} 
\centering
\captionof{table}{
    Slots parametrization.
    Best result in \textbf{bold}.
}
\label{tab:metadata:coordinate}
\resizebox{.95\textwidth}{!}{
\begin{tabular}{lcc}
   \toprule
Slot features & \gls{mse} $\downarrow$ & \gls{ndcg} $\uparrow$ \\
   \midrule
$x_c$, $y_c$                                   & $\bm{2.94 \pm 0.03}$ & $\bm{95.89 \pm 0.01}$ \\
$x_{\min}$, $x_{\max}$, $y_{\min}$, $y_{\max}$ & $3.03 \pm 0.02$ & $95.61 \pm 0.01$ \\
$x_{\min}$, $y_{\min}$, $w$, $h$               & $3.03 \pm 0.02$ & $95.59 \pm 0.02$ \\
   \bottomrule

\end{tabular}}
\end{minipage}
\hspace{0.005\linewidth}
\begin{minipage}[t]{0.43\linewidth}
\vspace*{0pt} 
\captionof{table}{
    Slot feature analysis.
    Best results in \textbf{bold}.
}
\label{tab:metadata:metadata}
\resizebox{.95\textwidth}{!}{
\begin{tabular}{cccc}
    \toprule
\multicolumn{2}{c}{Slot Metadata} &  &  \\  \cmidrule(lr){1-2} 
($x_c$, $y_c$) & type & \gls{mse} $\downarrow$ & \gls{ndcg} $\uparrow$ \\
    \midrule
$\checkmark$ & $\checkmark$ & $\bm{2.94 \pm 0.03}$ & $\bm{95.89 \pm 0.01}$ \\
 & $\checkmark$ &$3.06 \pm 0.02$ & $95.21 \pm 0.03$ \\
$\checkmark$ &  &$3.19 \pm 0.03$ & $95.04 \pm 0.03$ \\
    \bottomrule
\end{tabular}}
\end{minipage}
\end{table}

\Cref{tab:metadata:metadata} summarizes the feature importance of the slot metadata,
determined by the assessment of the \gls{mse} and \gls{ndcg} degradation when individual features are excluded. Removing the slot type leads to the most pronounced drop in performance metrics. Moreover , we verified that the order in which slots are fed into the AdSight model does not impact the quality of the results. Specifically, we compared the mean and standard deviation obtained for the proposed order 
(direct-top, direct-bottom, organic-top, organic-bottom) 
with those from two additional random orders. 
The results are identical, with $2.9 \pm 0.03$ for \gls{mse}
and $95.8 \pm 0.02$ for \gls{ndcg}.

\subsubsection{Cursor feature impact}
\label{sec:results:regression:cursor_impact}

\Cref{tab:cursor:feature_importance} (top part) provides an estimation of the importance of cursor features,
excluding the $2D$ coordinates. 
Feature importance is assessed by measuring the degradation in \gls{mse} and \gls{ndcg} when each feature is individually removed. 
Notably, the normalized sequence index emerges as the most critical feature, 
with its removal causing the most substantial deterioration in both \gls{mse} and \gls{ndcg}. 
Furthermore, \Cref{tab:cursor:feature_importance} (bottom part), reports the results of an ablation study of cursor features. 
Here, we incrementally remove features in decreasing order of importance, 
as indicated by the first four rows (top part) of \Cref{tab:cursor:feature_importance}. 
All comparisons were statistically significant ($p < .05$), except for the ablation study involving the removal of the index sequence feature.

\begin{table}[!ht]
\caption{
    Cursor features analysis. 
    Best results in \textbf{bold}.
}
\label{tab:cursor:feature_importance}
\centering
\resizebox{.95\linewidth}{!}{
\begin{tabular}{ccccccc}
    \toprule
 & \multicolumn{4}{c}{Cursor Features} &  &  \\  \cmidrule(lr){2-5} 
 & ($x$, $y$) & seq. index & pos. type & time & \gls{mse} $\downarrow$ & \gls{ndcg} $\uparrow$ \\
    \midrule
\multirow{4}{*}{\rotatebox{90}{Feature}} 
\multirow{4}{*}{\rotatebox{90}{Importance}}
& $\checkmark$ & $\checkmark$ & $\checkmark$ & $\checkmark$ & $\bm{2.94 \pm 0.03}$ & $\bm{95.89 \pm 0.01}$ \\
& $\checkmark$ & $\checkmark$ & $\checkmark$ &  & $3.05 \pm 0.03$ & $\bm{95.89 \pm 0.01}$ \\
& $\checkmark$ & $\checkmark$ &  & $\checkmark$ & $3.10 \pm 0.04$ & $95.87 \pm 0.01$ \\
& $\checkmark$ &  & $\checkmark$ & $\checkmark$ & $3.16 \pm 0.03$ & $95.86 \pm 0.01$ \\
    \midrule
\multirow{4}{*}{\rotatebox{90}{Feature}} 
\multirow{4}{*}{\rotatebox{90}{Ablation}}
& $\checkmark$ & $\checkmark$ & $\checkmark$ & $\checkmark$ & $\bm{2.94 \pm 0.03}$ & $\bm{95.89 \pm 0.01}$ \\
& $\checkmark$ &              & $\checkmark$ & $\checkmark$ & $3.16 \pm 0.03$ & $95.86 \pm 0.01$ \\
& $\checkmark$ &              &              & $\checkmark$ & $3.21 \pm 0.04$ & $95.83 \pm 0.02$ \\
& $\checkmark$ &              &              &              & $3.25 \pm 0.06$ & $95.75 \pm 0.03$ \\
    \bottomrule
\end{tabular}}
\end{table}

\begin{table}[b!]
\begin{minipage}[t]{0.44\linewidth}
\vspace*{0pt} 
\centering
\captionof{table}{
    Impact of the number of auxiliary slots $\bm{N}$.
    Best results in \textbf{bold}.
}
\label{tab:auxiliary_ads:number}
\resizebox{.95\textwidth}{!}{
\begin{tabular}{ccc}
    \toprule
$N$ & \gls{mse} $\downarrow$ & \gls{ndcg} $\uparrow$ \\
    \midrule
$4$ & $3.10 \pm 0.04$ & $95.79 \pm 0.02$ \\
$\bm{3}$ & $\bm{2.94 \pm 0.03}$ & $\bm{95.89 \pm 0.01}$ \\
$2$ & $3.05 \pm 0.02$ & $95.80 \pm 0.02$ \\
$1$ & $3.16 \pm 0.04$ & $95.77 \pm 0.02$ \\
$0$ & $3.22 \pm 0.05$ & $95.69 \pm 0.03$ \\
    \bottomrule
\end{tabular}}
\end{minipage}
\hspace{0.005\linewidth}
\begin{minipage}[t]{0.5\linewidth}
\vspace*{0pt} 
\captionof{table}{
    Impact of the weight $\bm{\alpha}$ 
    of the auxiliary loss. 
    Best results in \textbf{bold}.
}
\label{tab:auxiliary_ads:alpha}
\resizebox{.95\textwidth}{!}{
\begin{tabular}{cccc}
    \toprule
$N$ & $\alpha$ & \gls{mse} $\downarrow$ & \gls{ndcg} $\uparrow$ \\
    \midrule
$3$ & $0.00$ & $2.94 \pm 0.03$ & $95.89 \pm 0.01$ \\
$3$ & $\bm{0.33}$ & $\bm{2.86 \pm 0.02}$ & $\bm{96.07 \pm 0.02}$ \\
$3$ & $0.66$ & $2.88 \pm 0.02$ & $95.98 \pm 0.02$ \\
$3$ & $1.00$ & $3.16 \pm 0.04$ & $95.87 \pm 0.03$ \\
    \bottomrule
\end{tabular}}
\end{minipage}
\end{table}

\subsubsection{Auxiliary Slots Impact Study}
\label{sec:results:auxiliary_ads_impact}

To assess the contribution of auxiliary slots on model performance, we conducted two studies. In the first study, we set the $\alpha$ hyperparameter to zero, effectively disabling the influence of auxiliary prediction errors during training. By doing so, we isolate the impact of detailed cursor position categorization on model performance by means of excluding any learning from the auxiliary loss function. 
We then evaluated how performance varies as the number of auxiliary slots increases from $0$ to $4$. The results in \Cref{tab:auxiliary_ads:number} indicate that the inclusion of auxiliary slots consistently enhances performance ($N > 0$ outperforms $N = 0$), with the optimal performance observed at $N = 3$.

In the second study, we fixed $N=3$ and explored the role of the auxiliary loss function in the training process. 
The results, shown in \Cref{tab:auxiliary_ads:alpha}, 
demonstrate the effect of the auxiliary loss hyperparameter $\alpha$. Here, $\alpha = 0$ corresponds to excluding the auxiliary loss function, while $\alpha = 1$ assigns equal weight to the auxiliary loss and the standard slot loss. 
Our findings suggest that assigning a moderate weight to the auxiliary loss ($\alpha = 0.33$) yields the best performance.

\subsection{Classification task}
\label{sec:results:classification}

Similarly to the regression task, 
we evaluate AdSight (\gls{seq2seq} readout) 
against baseline models (\gls{mlp} readout) 
while varying both the cursor data representation and cursor embedding methods. 
In addition, we compare against baseline models~\cite{Arapakis20_mtdl}: 
BiLSTM, ResNet50, ResNet50-colour, and ResNet50-heatmap. 
These models use different representations: 
Time series, 
Colour trajectories, 
Colour trajectories and slot-specific colour, 
and Heatmap, respectively.

\Cref{tab:binary} summarises the classification results. 
Average Performance is calculated using weighted averages, 
based on the distribution of slot categories within the datasets. 
Specifically, the \gls{seq2seq} approach consistently outperforms the \gls{mlp} baseline 
across all configurations of cursor data representation and embeddings. 
Additionally, employing a time-series cursor data representation along with a Transformer encoder-based cursor embedding 
yields optimal average results, irrespective of the readout mechanism. 
Furthermore, the \gls{seq2seq} model, combined with a time-series cursor data representation and Transformer encoder-based embedding, 
achieves the best average performance.

The models utilizing a time-series cursor data representation consistently excel in the classification tasks 
for direct-top and direct-right slot categories, 
while models leveraging a visual representation of cursor data demonstrate superior performance 
in the organic-bottom classification task. 
Performances on the organic-top classification task 
are comparable between models using time-series cursor data representation 
and those employing visual representations, 
with the former slightly outperforming the latter. 

Last, the baseline models proposed by \citet{Arapakis20_mtdl} 
achieve performance comparable to our baselines using \gls{mlp} as the readout mechanism 
but fall short as compared to \gls{seq2seq} models. 
The Wilcoxon signed-rank test (with correction for multiple testing) revealed that
models with \gls{seq2seq} readout consistently outperformed those with \gls{mlp} readout
across all slot-type/metric/embedding combinations ($p < .05$). 
For direct-top/right and organic-top slots, 
the combination of \gls{seq2seq} readout and Transformer embedding
was statistically superior to the second-best model ($p < .05$). 
For organic-bottom slots, 
models with \gls{seq2seq} readout using visual cursor data representation performed similarly, 
with the best model outperforming those using time-series cursor data ($p < .05$). 
These findings demonstrate the robustness and superior accuracy of the \gls{seq2seq} approach, 
particularly when paired with a time-series representation and Transformer-based embeddings.

\section{Conclusion\label{sec:discussion_and_future_work}}

We introduced AdSight, a novel Transformer-based Seq2Seq model to predict user attention in multi-slot environments such as \glspl{serp}. 
By leveraging mouse cursor trajectories and slot-specific metadata, 
AdSight achieved robust performance in both regression and classification tasks, 
delivering an accurate quantification of user attention.
Our findings highlight the importance of incorporating slot metadata and auxiliary slots.
Furthermore, the Seq2Seq architecture outperformed traditional MLP-based baselines across all tested configurations, 
demonstrating its effectiveness in capturing complex intra- and inter-sequence relationships, 
and paving the way for low-cost scalable inference of user attention.

\begin{acks}
This research is supported by the Pathfinder program of the European Innovation Council (SYMBIOTIK project, grant 101071147).
\end{acks}

\end{document}